%Paper: hep-th/9301023
%From: Tsukasa Tada (YITP, Uji) <tada@yisun1.yukawa.kyoto-u.ac.jp>
%Date: Fri, 08 Jan 1993 20:24:53 +0900
%Date (revised): Sat, 16 Jan 1993 11:35:43 +0900

%%USE phyzzx and psfig%%%%%%%%%%%%%%%%%%%%%%%%%%%%%%%%%%%%%%%%%%%%%%%
%%
%%In case psfig is not available, please comment out the corresponding
%%lines.
%%Here we have 8 figures, attached at the end of the file
%%as tar compressed uuencoded form
%%
\input phyzzx
\input psfig
%%%%% YITP, Uji macros
\def\unlock{\catcode`@=11} % This allows us to modify PLAIN macros.
\def\lock{\catcode`@=12} % at signs are no longer letters
\unlock
%%%%%%%%%%%%%%%%%%%%%%%%%%%%%%%%%%%%%%%%%%%%%%%%%%%%%%%
%% First comes the CORRECTIONS of PHYZZX(Y).TEX macro
%%%%%%%%%%%%%%%%%%%%%%%%%%%%%%%%%%%%%%%%%%%%%%%%%%%%%%
\paperfootline={\hss\iffrontpage\else\ifp@genum\tenrm
    -- \folio\ --\hss\fi\fi}
\def\titlestyle#1{\par\begingroup \titleparagraphs
     \iftwelv@\fourteenpoint\fourteenbf\else\twelvepoint\twelvebf\fi
   \noindent #1\par\endgroup }
\def\GENITEM#1;#2{\par \hangafter=0 \hangindent=#1
    \Textindent{#2}\ignorespaces}
\def\papersize{\hsize=35pc \vsize=52pc \hoffset=0.5pc \voffset=0.8pc
   \advance\hoffset by\HOFFSET \advance\voffset by\VOFFSET
   \pagebottomfiller=0pc
   \skip\footins=\bigskipamount \normalspace }
\papers  %  This is the default
\def\address#1{\par\kern 5pt \titlestyle{\twelvepoint\sl #1}}
\def\abstract{\par\dimen@=\prevdepth \hrule height\z@ \prevdepth=\dimen@
   \vskip\frontpageskip\centerline{\fourteencp Abstract}\vskip\headskip }
\newif\ifYITP \YITPtrue
\font\fourteenmib =cmmib10 scaled\magstep2    \skewchar\fourteenmib='177
\font\elevenmib   =cmmib10 scaled\magstephalf   \skewchar\elevenmib='177
\def\YITPmark{\hbox{\fourteenmib YITP\hskip0.2cm
        \elevenmib Uji\hskip0.15cm Research\hskip0.15cm Center\hfill}}
\def\titlepage{\FRONTPAGE\papers\ifPhysRev\PH@SR@V\fi
    \ifYITP\null\vskip-1.70cm\YITPmark\vskip0.6cm\fi %this line is added
   \ifp@bblock\p@bblock \else\hrule height\z@ \rel@x \fi }
%%%%%%%%%%%%%%%%%%%%%%%%%%%%%%%%%%%%%%%%%%%%%%%%
%% end of the CORRECTIONS of PHYZZX(YU).TEX  %%%
%%%%%%%%%%%%%%%%%%%%%%%%%%%%%%%%%%%%%%%%%%%%%%%%
%%%%%%%%%%%%%%%%%%%%%%%
\def\schapter#1{\par \penalty-300 \vskip\chapterskip
   \spacecheck\chapterminspace
   \chapterreset \titlestyle{\ifcn@@\S\ \chapterlabel.~\fi #1}
   \nobreak\vskip\headskip \penalty 30000
   {\pr@tect\wlog{\string\chapter\space \chapterlabel}} }

%%%%%%%%%%%%%%%%%%%%%%%%%
\def\ssection#1{\par \ifnum\lastpenalty=30000\else
   \penalty-200\vskip\sectionskip \spacecheck\sectionminspace\fi
   \gl@bal\advance\sectionnumber by 1
   {\pr@tect
   \xdef\sectionlabel{\ifcn@@ \chapterlabel.\fi
       \the\sectionstyle{\the\sectionnumber}}%
   \wlog{\string\section\space \sectionlabel}}%
   \noindent {\S \caps\thinspace\sectionlabel.~~#1}\par
   \nobreak\vskip\headskip \penalty 30000 }
%%%%%%%%%%%%%%%%%%%%%%%%%%%%%%%%%%%%%%%%%%%%%%%%%%%%%%
%%%%  YITP Uji Res. Ctr. preprint. header part is above
%%%%%%%%%%%%%%%%%%%%%%%%%%%%%%%%%%%%%%%%%%%%%%%%%%%%%%
\def\YITP{\address{Uji Research Center \break
               Yukawa Institute for Theoretical Physics\break
               Kyoto University,~Uji 611,~Japan}}
\newtoks\pubnum
\Pubnum={YITP/U-\the\pubnum}
\pubnum={9?-??}
%%%%%%%%%%%%%%%%%%%%%%%%%%%%%%%%%%%%%%

%

%

%
\def\globaleqnumbers{\relax\if\equanumber<0\else\global\equanumber=-1\fi}
%
%%%%% Add equation number by 1
\def\addeqno{\ifnum\equanumber<0 \global\advance\equanumber by -1
    \else \global\advance\equanumber by 1\fi}
%%%%%%%%%%%%%%%%%%%%%%%%%%%
\mathchardef\Lag="724C  %%%% Lagrangian symbol = \Lag

 %%  definition of \yen mark
%
%%End equation mark with equation number.
 
%%Draw a box around a thing.
 
%%Figure mark in text
 
%%Table mark in text
 
%%Over Bar ---> Use \overline
 \def\overbar#1{\vbox{\ialign{##\crcr
           \vrule depth 2mm
           \hrulefill\vrule depth 2mm
           \crcr\noalign{\kern-1pt\vskip0.125cm\nointerlineskip}
           $\hfil\displaystyle{#1}\hfil$\crcr}}}
%% Enforce carriage return and line feed within a paragraph

%%rho with a little bit up

%
 %Equation number as you like
%
%A little bigger cdot

%%%%%%%%%%%
%square (in math mode)
%
\def\sqr#1#2{{\vcenter{\hrule height.#2pt
      \hbox{\vrule width.#2pt height#1pt \kern#1pt
          \vrule width.#2pt}
      \hrule height.#2pt}}}

%%%%%%%%%

\def\Buildrel#1\under#2{\mathrel{\mathop{#2}\limits_{#1}}}
\def\llongrarrow{\hbox to 40pt{\rightarrowfill}}

%%%%%%%%%%
\def\journals#1&#2(#3){\unskip; \sl #1~\bf #2 \rm (19#3) }
%%%%%%%%%%%%%%%%%%%%%%%%%%%%%%%%%%%
%% item number at the leftmost end
%
 \def\nllap#1{\hbox to-0.35em{\hskip-\hangindent#1\hss}}

%%%%%%%%%%%%%%%%%%%%%%

%
\def\rslash{\partial\kern-0.026em\raise0.17ex\llap{/}%
          \kern0.026em\relax}
\def\Dslash{D\kern-0.15em\raise0.17ex\llap{/}\kern0.15em\relax}
%%%%%%%%%%%%%%%%%%%%%%%
\mathchardef\bigtilde="0365

%%%%%%%%%%%%%%%%%%%%%%%%%%%%%%%%%%%%%%%%
%   double eqalign macro (\deqalign{...})
%
\def\deqalign#1{\null\,\vcenter{\openup1\jot \m@th
    \ialign{\strut\hfil$\displaystyle{##}$&$\displaystyle{##}$&$
	\displaystyle{{}##}$\hfil\crcr#1\crcr}}\,}
%%%%%%%%%%%%%%%%%%%%%%%%%%%%%%%%%%%%%%%%%%%%%%%%%%%%%%%%%%
%
% [ \seqalignno{..&..&\seq\cr} \eqs ]
%
%  "\seqalignno" gives alignned equations with eq-numbers like
%     (1.2a),(1.2b) ... with "\seqn{...}" and/or "\seq"[=\seqn\?].
%  "\eqsname{...}" or "\eqs"[=\eqsname\?] gives an eq-number
%     like (1.2) to the alignned equations (1.2a),(1.2b)... .
%     ("\seqn" can be used also in \eqalignno: \seqn=\eqnalign.)
%
\newcount\eqabcno \eqabcno=97 %% \char97=``a''
\newcount\sequanumber \sequanumber=0
\newif\iffirstseq \firstseqtrue
\newif\ifequationsabc \equationsabcfalse
\def\eqnameabc#1{\relax\pr@tect
    \iffirstseq\global\firstseqfalse%
        \ifnum\equanumber<0 \global\sequanumber=\number-\equanumber
           \xdef#1{{\rm(\number-\equanumber a)}}%
           \global\advance\equanumber by -1
        \else \global\advance\equanumber by 1
           \xdef#1{{\rm(\ifcn@@ \chapterlabel.\fi \number\equanumber a)}}
        \fi
    \else\global\advance\eqabcno by 1
        \ifnum\equanumber<0
           \def#1{{\rm(\number\sequanumber \char\number\eqabcno)}}%
        \else
           \xdef#1{{\rm(\ifcn@@ \chapterlabel.\fi%
                        \number\equanumber \char\number\eqabcno)}}
        \fi
    \fi}

\def\eqsname#1{\relax\pr@tect%
        \ifnum\equanumber<0
           \def#1{{\rm(\number\sequanumber)}}
        \else
           \xdef#1{{\rm(\ifcn@@ \chapterlabel.\fi \number\equanumber)}}
        \fi}

%%%%%%%%%%%%%%%%%%%%%%%%%%%%%%%%%%%%%%%%%%%%%%%%%%%%%%%%%%%%%%%%%%%%%%%
%% \draft causes the symbolic names of equations to be printed
%% alongside the equation numbers

%%%%%%%%%%%%%%%%%%%%%%%%
  % to save previous macro.
%
\lock
%%%%%%%Here starts the content
%%%%%%%
\pubnum={92-38}
\date={December 1992}
\titlepage
\vskip0.5cm
\title{Reflection equation and link polynomials \break
for arbitrary genus solid tori}
\author{C. Schwiebert\footnote{*}{Feodor Lynen and JSPS Fellow.
Supported in part by Grant in aid from Ministry of Education, Science
and Culture, No.91081. Address from January 1993: Institute for
Theoretical Physics,University of Karlsruhe,Germany}}
\YITP
\abstract{The correspondence between the braid group on a solid torus of
arbitrary genus and the algebra of Yang-Baxter and reflection equation
operators is shown. A representation of this braid group in terms of
$R$-matrices is given. The characteristic equation of the reflection
equation matrix is considered as an additional skein relation. This
could lead to an intrinsic definition of invariant link polynomials on
solid tori and, via Heegaard splitting, to invariant link polynomials
on arbitrary three-manifolds without boundary.}

%%%%%%%%%%%%%%%%%%%%%%%%%  References  %%%%%%%%%%%%%%%%%%%%%%%%%%%%%%%%%%%%%%%

\def \ADW {Y. Akutsu, T. Deguchi and M. Wadati,
J. Phys. Soc. Jpn. {\bf 56} (1987) 839, 3039, 3464;
{\bf 57} (1988) 757, 1173, 1905}
\def \Alex {J. W. Alexander, Trans. Am. Math. Soc. {\bf 20} (1923) 275;
Proc. Natl. Acad. Sci. {\bf 9} (1928) 93}
\def \Art {E. Artin, Ann. Math. {\bf 48} (1947) 101}
\def \CWSSW {U. Carow-Watamura, M. Schlieker, M. Scholl and  S. Watamura,
Z. Phys. C {\bf 48} (1990) 159}
\def \Con {J. H. Conway, in \lq\lq Computational problems in abstract
algebra \rq\rq, Pergamon, New York 1970}
\def \Dria {V. G. Drinfeld, in: Proc. ICM-86, {\bf 1}, Berkeley,
AMS (1987) 798;
and in: \lq\lq Yang-Baxter equation in integrable systems\rq\rq,
ed. by M. Jimbo, World Scientific, Singapore 1990}
\def \FRT {L. D. Faddeev, N. Yu. Reshetikhin and L. A. Takhtajan, Alg. i Anal.
{\bf 1} (1989) 178 (in Russian, English transl.: Leningrad Math. J.
{\bf 1} (1990) 193)}
\def \Jim {M. Jimbo, Lett. Math. Phys. {\bf 10} (1985) 63;
{\bf 11} (1986) 247;
Commun. Math. Phys. {\bf 102} (1986) 537}
\def \Jon {V. F. R. Jones, Bull. Am. Math. Soc. {\bf 12} (1985) 103;
Ann. Math. {\bf 126} (1987) 335}
\def \Kauf {L. H. Kauffman, Topology {\bf 26} (1987) 395}

\def \Koh {\lq\lq New developments in the theory of knots\rq\rq,
ed. by T. Kohno, World Scientific, Singapore 1989}
\def \KSS {P. P. Kulish, R. Sasaki and C. Schwiebert, Kyoto Univ. preprint
YITP/U-92-07, March 1992; to appear in J. Math. Phys.}
\def \Mada {S. Majid, J. Math. Phys. {\bf 32} (1991) 3246}
\def \Madb {S. Majid, Cambridge Univ. preprint DAMTP/92-12, February 1992}
\def \Man {Yu. I. Manin, Montreal Univ. preprint CRM-1561, 1988}
\def \Pod {P. Podles, Lett. Math. Phys. {\bf 14} (1987) 193}
\def \Sos {A. B. Sossinsky, \  in:
\lq\lq  Euler int. math. inst. on quantum groups\rq\rq, \  ed. by \nextline
P. P. Kulish, Lect. Notes Math. {\bf 1510}, Springer, Berlin 1992}
\def \MoRe {G. Moore and N. Yu. Reshetikhin, Nucl. Phys. {\bf B328}
(1989) 557}
\def \SmWZ {W. B. Schmidke, J. Wess and B. Zumino, Z. Phys. C {\bf 52}
(1991) 471}
\def \Sklb {P. P. Kulish and E. K. Sklyanin, Kyoto Univ. preprint
YITP/K-980, May 1992; to appear in J. Phys. A}
\def \Tur {V. G. Turaev, Invent. Math. {\bf 92} (1988) 527}
\def \Wit {E. Witten, Commun. Math. Phys. {\bf 121} (1989) 351}
\def \WZ {J. Wess and B. Zumino, Nucl. Phys. (Proc. Suppl.)
{\bf B18} (1990) 302}

%%%%%%%%%%%%%%%%%%%   Abbreviations   %%%%%%%%%%%%%%%%%%%%%%%%%%%%%%%%%%%%

\def\Sl#1{$SL_q(#1)$}
\def\sl#1{$sl_q(#1)$}
\def\Rt{\ifmmode {\widetilde R} \else {$\widetilde R$} \fi}
\def\Rh{\ifmmode {\widehat R} \else {$\widehat R$} \fi}
\def\Rth{\ifmmode {\widehat {\widetilde R}} \else
                  {$\widehat {\widetilde R}$} \fi}
\def\a{\alpha}
\def\b{\beta}
\def\c{\gamma}
\def\d{\delta}
\def\e{\varepsilon}
\def\s{\sigma}
\def\t{\tau}
\def\w{\omega}
\def\NPrefs{\let\refmark=\NPrefmark}
\NPrefs

%%%%%%%%%%%%%%%%%%%%%&    Section 1   %%%%%%%%%%%%%%%%%%%%%%%%%%%%%%%%%%
\section{Introduction}
In this article we should like to point out a relation between the
reflection equation (RE) and the braid group on a 3-manifold of arbitrary
genus with boundary.

We shall begin by introducing the RE as a quantum group (QG) comodule,
then derive its properties as a
non-commutative associative algebra and mention several applications
of the RE. Concerning the braid group we start from the results of A.
B. Sossinsky who defined the braid group on a solid handlebody of
arbitrary genus. It consists of the usual generators for genus zero
3-manifolds and additional ones implementing windings around handles.
We show that the relations they obey are precisely the Yang-Baxter
equation (YBE) and the RE. We find an explicit representation of this
algebra in terms of $R$-matrices and quantum algebra generators.
We derive quadratic characteristic
equations for the additional generators and suggest their
interpretation as new skein relations. In principle this defines
invariant link polynomials for closed braids in
arbitrary genus 3-manifolds with boundary. This in turn, via Heegaard
splitting, might pave the way for constructing invariant polynomials
for links in arbitrary 3-manifolds without boundary.

%%%%%%%%%%%%%%%%%%%%%%%%%  Section 2   %%%%%%%%%%%%%%%%%%%%%%%%%%%%

\section{The reflection equation}

In this section we introduce the reflection equation, discuss its
properties and point out some applications. Many of them were
considered earlier in \Ref\rKSS{\KSS}, but some have been obtained
only afterwards. We will not say anything about one-dimensional
representations of the RE algebra \refmark{\rKSS} as for the braid
group at least two-dimensional representations are needed.
Furthermore, throughout this paper we only discuss the standard
example of \sl2. We assume familiarity of the reader with this and
basic quantum group terminology as introduced in \Ref\rFRT{\FRT} for
example.

We start from scratch and ask for the possibility of generalizing
statistics of (space-time) coordinates, i.e. introduce 2-`spinors'
$ x^i=\pmatrix{u \cr v \cr}, i=1,2 $ with commutation relations
$$ u v = q v u ,    \eqn \spin  $$
where $q \in {\bf C} $ instead of $q = 1$ for bosonic or $q = -1$ for
fermionic coordinates. We set up a covariant notation and rewrite
\spin\  as
$$ x^i x^j = q^{-1} {R^{ji}}_{kl} x^k x^l ,  \eqn \manx  $$
where we have to introduce the $R$-matrix ($(ji)$ labeling columns
and $(kl)$ rows with natural index order $(11,12,21,22)$; vanishing
entries are omitted)
$$ {R^{ji}}_{kl} = \pmatrix{q& & & \cr  &1& & \cr  &\w&1& \cr  & & &q\cr}
, \quad
\w=q-q\sp {-1}  \eqn\rmat $$
and for consistency of higher order relations it has to satisfy the
Yang-Baxter equation
$$ R_{12} R_{13} R_{23} = R_{23} R_{13} R_{12} .  \eqn \ybe $$
Here it was possible to hide all indices by introducing the standard
matrix notation $ R_{12} = R \otimes I $ etc., with $I$ the identity
matrix, acting in a triple product
of vector spaces \refmark{\rFRT}. Above construction is just the
quantum plane as introduced in \Ref\rMan{\Man} and generalized in
\Ref\rWZ{\WZ}.

We define a transformation of $x^i$ as
$$ {x'}^i = {T^i}_j x^j , \eqn \trx  $$
and ask what restrictions on the $ {T^i}_j $ result if we demand
invariance of the basic relation \manx. The outcome is the celebrated
QG equation
$$ R T_1 T_2 = T_2 T_1 R , \eqn \rtt  $$
and again it was  possible to get rid of indices by using the
notation $ T_1 = T \otimes I $, $ T_2 = I \otimes T $.
We do not bother to write out \rtt\  for the entries of $T$ as we will
not need it. The QG \Sl2 is defined by \rtt\  if we set equal to one the
quadratic central element corresponding to the generalized
determinant. In fact, this is a non-commutative and non-cocommutative
Hopf algebra, not a Lie group but a deformation of $SL(2)$
\REF\rDria{\Dria} \REF\rJim{\Jim} \refmark{\rFRT,\rDria,\rJim}.

We further introduce a second quantum plane $ y_i $, however with
lower index and transforming by $ T^{-1} $ (the antipode of $T$)
$$ {y'}_i = y_j {(T^{-1})^j}_i .  \eqn \try  $$
The commutation relations of $y_i$ are completely fixed, using an
ansatz $ y_i y_j = \a y_k y_l {M^{kl}}_{ij} $, $\a =\  $const., we
transform according to \try\  and use \rtt\  to get
$$  y_i y_j = q^{-1} y_k y_l {R^{kl}}_{ji} .  \eqn \many $$
We can even calculate the commutation relations between $x^i$ and
$y_j$ this way
$$  x^i y_j = \a y_k x^l {R^{ik}}_{lj} , \eqn \manxy  $$
this time however the constant $\a$ is not fixed by covariance, we do
not need it and may set $\a = 1$. The inverse of \manxy\  is
$$  y_j x^i = {\a}^{-1} {(((R^{t_2})^{-1})^{t_2})^{il}}_{kj} x^k y_l ,
\eqn \manyx  $$
where the superscript $t_2$ means transposition in the second space in
which $R$ acts, i.e. interchange of $(i \leftrightarrow l)$ in \rmat.

Next we consider a product of the two quantum planes and define the
matrix $ {K^i}_j = x^i y_j $. It transpires that $K$ transforms as $ K'
= T K T^{-1} $, i.e. multiplying together quantum planes we can
construct tensors of arbitrary rank covariant w.r.t. the QG coaction.
They can straightforwardly be $q$-(anti)symmetrized in analogy to the
non-deformed case. Again, the commutation relations of ${K^i}_j$ are
completely determined by \manx, \many, \manxy\  and \manyx
$$  {K^i}_j {K^m}_n = q^{-2} {(((R^{t_2})^{-1})^{t_2})^{ml}}_{kj}
{R^{ki}}_{k'l'} {R^{rs}}_{nl} {R^{l'r'}}_{s'r} {K^{k'}}_{r'}{K^{s'}}_s
. \eqn \kcom  $$
Hearty readers may check that the entries of $ K = \pmatrix{a&b\cr
c&d\cr} $ satisfy the commutation relations
$$ \eqalign{ab&=q\sp {-2}ba , \cr  ac&=q\sp 2ca , \cr}  \qquad
\eqalign{ad&=da , \cr  bc-cb&=q\sp {-1}\w (ad-a\sp 2) , \cr}  \qquad
\eqalign{bd-db&=-q\sp {-1}\w ab , \cr  cd-dc&=q\sp {-1}\w ca  .
\cr} \eqn \alg  $$
This algebra has two central elements, the quantum trace and the
quantum determinant which we set equal to one
$$ c_1 = a + q^2 d , \qquad  c_2 = a d - q^2 c b \equiv 1 . \eqn \cent
$$
The algebra depends only on $q^2$ and one may rescale $q^2 \rightarrow
{\widetilde q}$, then if ${\widetilde q}$ is a phase,
${\widetilde q}^N = 1$, we find that
$a^N$ is a further central element. \nextline%

We can define the `antipode' $S(K) \equiv K^{-1}$ as
$$ \eqalign{S(a) &= q^2 d - q \w a , \cr S(c) &= - q^2 c , \cr} \qquad
   \eqalign{S(b) &= - q^2 b , \cr S(d) &= a . \cr} \eqn \antip  $$
Then we easily establish a relation (characteristic equation) between
$K$ and $K^{-1}$
$$ K^{-1} = -q^2 K + c_1 I , \eqn \cheq  $$
which will play the role of a skein relation later on. \nextline%pen

If we impose suitable reality conditions on $x^i , y_j $ and hence $
{K^i}_j $ then a linear combination of the elements of \alg\  is just the
$q$-deformed Minkowski space
\REF\rCWSSW{\CWSSW} \REF\rSmWZ{\SmWZ} \refmark{\rCWSSW,\rSmWZ}, where
$c_1$ is the time coordinate and $c_2$ the invariant length. Various
reality conditions are discussed in \Ref\rSklb{\Sklb}. \nextline%
Truncation of algebra \alg\  by $c_1 = 0$ can be shown to lead to the
quantum 2-sphere of Podles, a quantum analogue of homogeneous spaces
\Ref\rPod{\Pod}. \nextline%

Notice also the similarity of \kcom\  to the basic relation \manx\  if one
reads the four $R$-matrices on the RHS as a single one with four pairs
of indices \refmark{\rCWSSW}. It is simple to introduce an index free
notation for quantum planes and extend it to the $K$-matrix, but we
shall not go into this here \REF\rMadb{\Madb} \refmark{\rSklb,\rMadb}.

A basic observation is now that relations \alg\  can be encoded in a
matrix equation
$$ R K_1 \Rt K_2 = K_2 R K_1 \Rt , \eqn \re  $$
where $\Rt = P R P$ with $P$ the permutation operator. This is one of the
RE discussed in \refmark{\rKSS}. It is easy to establish invariance
of \re\  under the QG coaction $K' = T K T^{-1}$. The other RE discussed
extensively in \refmark{\rKSS} is invariant under $K' = T K T^t$,
where the superscript $t$  denotes the transpose of $T$. For the
special choice of \rmat\  as $R$-matrix its algebra is isomorphic to
\alg. We do not write it down here as it is irrelevant for our
purposes, but one should keep in mind that different types of QG
covariant tensors can be constructed.

It is remarkable that one can
define a twisted (or braided) `coproduct' of the same form as for
the QG, i.e. $\Delta (K) = K {\dot \otimes} K$, however with
non-commutativity between elements of different spaces. We avoid
tensor product notation and distinguish elements of different spaces
by a prime. So it is easy to prove the following: \nextline
Given two different solutions $K$ and $K'$ of \re\  then
$$ (i) \quad {\widetilde K} = K K' ,  \qquad  {\rm and} \qquad
(ii) \quad {\widetilde {\widetilde K}} = K K' K^{-1} \eqn \prod $$
are also solutions of \re\  provided $K$ and $K'$ commute as follows
$$  R K_1 R^{-1}K'_2 = K'_2 R K_1 R^{-1} . \eqn \retwi  $$
This gives 16 commutation relations between the elements of $K$ and
$K'$, we do not bother to show them explicitly here.
\nextline%

This process of building up new solutions can obviously be continued,
but some care has to be taken to keep track of the ordering and
multiplying from the correct side as \retwi\
is not symmetric under exchange of $K$ and $K'$. Due to \prod\  we may
interprete \alg\  as an algebra having a braided (adjoint) coaction on
QG comodules. Algebra \alg\  was also constructed in
\Ref\rMada{\Mada} from the point of view of braided tensor categories.
\nextline

An important point is that the central elements of $K$ and $K'$ are
mutually central in both algebras, i.e.
$$ [{K^i}_j,c'_m] = [{{K'}^i}_j,c_m] = 0 , \quad m=1,2 . \eqn \mutcen $$
\nextline
It is obvious that we have central elements for the combined solutions
and also characteristic equations, for example
$$ (KK')^{-1} = -q^2 KK' + C_1 I , \eqn \comcen $$
where $ C_1 = a a' + b c' + q^2 (c b' + d d') $.

A further property of the RE will be needed later on. The \sl2 algebra
dual to the QG \rtt\  can be written in matrix form as \refmark{\rFRT}
$$ \Rt L^{\e_1}_1 L^{\e_2}_2 = L^{\e_2}_2 L^{\e_1}_1 \Rt , \qquad
(\e_1,\e_2) = \{(+,+),(+,-),(-,-)\}  \eqn \rll  $$
where
$$ L^+ = \pmatrix{q^{H/2} & q^{-1/2} \w X^- \cr 0 & q^{-H/2} \cr} ,
\qquad L^- = \pmatrix{q^{-H/2} & 0 \cr -q^{1/2} \w X^+ & q^{H/2} \cr}
\eqn \lplm $$
and this gives the \sl2 algebra
$$ [H,X^{\pm}] = \pm 2 X^{\pm} , \qquad
   [X^+,X^-] = \w^{-1} (q^H - q^{-H})  \eqn \sltwo  $$
with antipode $ S(H) = - H $, $ S(X^{\pm}) = - q^{\mp 1} X^{\pm} $
and coproduct $ \Delta(L^{\pm}) = L^{\pm} {\dot \otimes} L^{\pm} $. It is
easy to show using \rll\  that $ K = S(L^-) L^+ $ is a solution of the
RE \Ref\rMoRe{\MoRe}, explicitly $K$ is given by
$$ K = \pmatrix{q^H & q^{-1/2} \w q^{H/2} X^- \cr
q^{-1/2} \w X^+ q^{H/2} & q^{-H} + q^{-1} \w^2 X^+ X^- \cr} .
\eqn \lsol $$
The RE algebra hence plays different roles, it is a comodule w.r.t.
the QG (with a `coaction' of it on QG comodules, see above)
and on the other hand it acts via \lsol\  on representations of the
quantum algebra dual to the QG. Formally, there is also an action of
the quantum algebra generators $ L^{\pm} $ on the RE algebra
\refmark{\rSklb,\rMada}. However,
interpretations of the RE algebra are different in each case. We refer
the reader to \refmark{\rKSS} for further applications of the RE  and
historical development, we are quite confident that still more
applications can be uncovered.

%%%%%%%%%%%%%%%%%%%%%%%    Section 3    %%%%%%%%%%%%%%%%%%%%%%%%%%%%%%%

\section{Braid group on solid tori}

The braid group $B^g_n$ in a solid handlebody $H_g$ of genus $g$ was
derived in \Ref\rSos{\Sos}. In addition to the generators $ \s_i ,
i=1,\ldots,n-1 $ of the braid group $B_n$ defined on a 3-dimensional
manifold of genus zero there are generators $ \t_\a , \a=1,\ldots,g $
implementing windings around the $g$ handles. The  algebra
is given as
$$\eqalign{\s_i \s_{i+1} \s_i &= \s_{i+1} \s_i \s_{i+1} , \quad
i=1,\ldots,n-1 \cr
           \s_i \s_j &= \s_j \s_i , \quad \mid i-j \mid \geq 2 \cr
           \s_i \t_\a &= \t_\a \s_i , \quad i \geq 2 , \ \a =1,\ldots,g
\cr            \s_1 \t_\a \s_1 \t_\a &= \t_\a \s_1 \t_\a \s_1 , \quad
\a =1,\ldots,g \cr
           \s_1 \t_\a \s_1^{-1} \t_\b &= \t_\b \s_1 \t_\a \s_1^{-1} ,
\quad \a < \b , \  \a,\b=1,\ldots,g \cr} \eqn \bralg $$
and the first two relations define the well known Artin braid group
\Ref\rArt{\Art}. This group $B^g_n$ is a subgroup of $B_{g+n}$ as
explained in \refmark{\rSos}. We refer to this paper for further
details and references. Here we only explain conventions\footnote*{We
found it necessary to have conventions slightly different from
\refmark{\rSos}, especially in the last formula of \bralg\  the
condition in \refmark{\rSos} is $ \b < \a $, and for lines leaving the
unit cube $ z_1 > z_2 $ (see text below).} briefly which should make
\bralg\  fairly transparent.

On the handlebody (Fig.1), without loss of generality, we prescribe a fixed
ordering of the points where the strands begin (resp. end) having
coordinates $ P_i^{(1)} = ({i \over {n+1}},{1 \over 2},1)$
\ (resp. $ P_j^{(0)} = ({j \over {n+1}},{1 \over 2},0)$,
\ $i,j=1,\ldots,n$ in a lefthanded $(x,y,z)$-coordinate system. So the
unit cube in the positive octant
is contained in $H_g$ and the usual braids are obtained by connecting
points $P_i^{(1)}$ and $P_j^{(0)}$ by strands confined to the unit
cube. The braid diagram is obtained by projecting on the $x$-$z$-plane.
The handles are positioned, say, to the left of the unit cube around
coordinates $ h_\a = ({-\a \over {g+1}},y,1-{\a \over {g+1}}), \
\a =1,\ldots,g $. For the braid group on $H_g$ the strands are allowed
to leave the unit cube at
height $z$ and go around the handle $h_\a $ counterclockwise for $\t_\a $
(clockwise for $\t_{\a}^{-1}$) and then come back to the unit cube at
height $z-\d$, $\d$ small. The convention is that strands leaving or
entering the unit cube at height $z_1$ should be under those doing so
at $z_2$ in the projection onto the $x$-$z$-plane if $z_2 > z_1$. Within
the unit cube strands can only go downward in the negative
$z$-direction. This definition can be further formalized, but all this
is rather intuitive.
\vskip1cm
\centerline{\psfig{figure=LINKfig1.ps,width=10cm}}
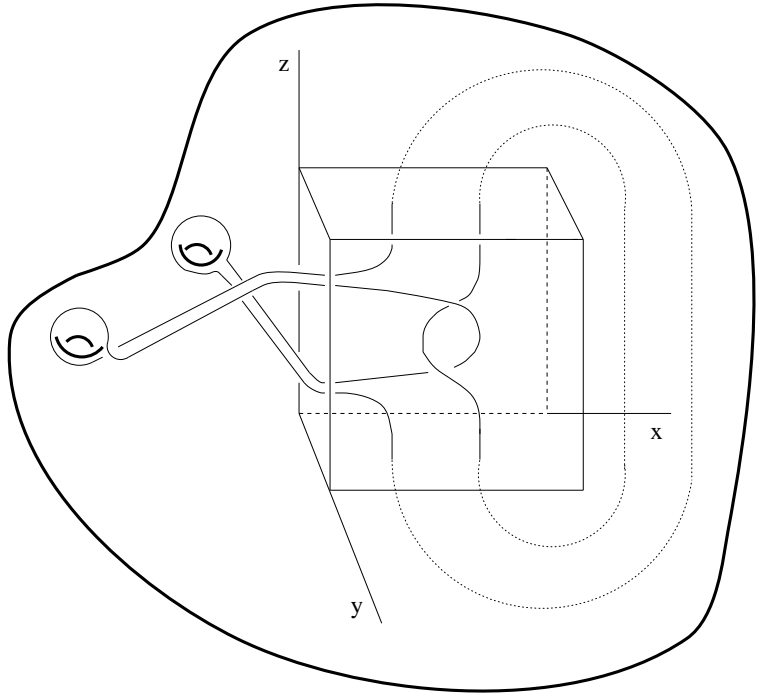
\centerline{Fig.1: \  The 2-braid $\t_2^{-1} \s_1^{-2} \t_1$ and its
closure (dotted lines)}
\vskip1cm
The strands leaving the unit cube always belong to the first space
$V_1$ of the tensor product $ V(n) = V_1 \otimes \ldots \otimes V_n$ on
which the $\s_i$ act, and this explains why only $\s_1$ is non-commuting
with $\t_\a $. For our arguments it is more appropriate to think of
piercing long needles through the handles and after that forget about
them. Then, if we rotate the needles by $\pi / 4$ around the $x$-axis
counterclockwise to $ h'_\a = ({-\a \over {g+1}},{\a \over {g+1}}-1,z) $
we can depict the braiding in a suggestive way by projecting on the
$x$-$z$-plane (Fig.2). All relations of \bralg\ and all those
involving $K$ like \re, \prod\  and \retwi\   can be represented
in terms of diagrams as in Fig.2 and they are proven easily this way.
\vskip1cm
\centerline{\psfig{figure=LINKfig2.ps,width=10cm}}
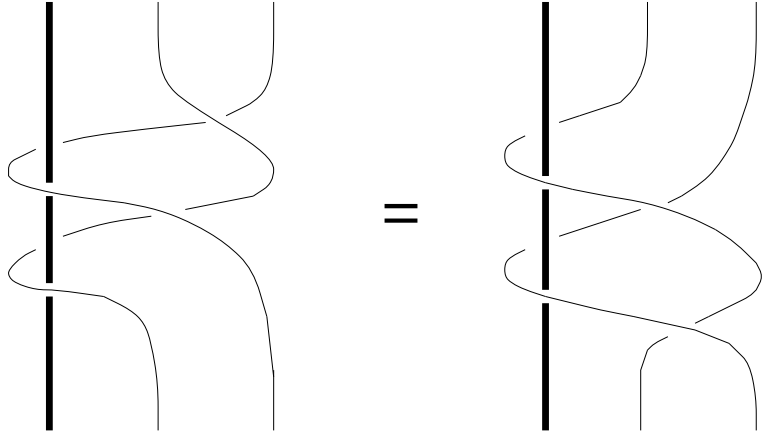
\centerline{Fig.2: \  A graphical representation of the reflection equation}
\vskip1cm
A $g$-link $L_g$ on $H_g$ is obtained by connecting $P^{(0)}_i$ with
$P^{(1)}_i$ outside the unit cube in the $ x > 0 $ region. Then, citing a
theorem \refmark{\rSos}, every $g$-link can be obtained as the closure
of a $g$-braid. However, the Markov theorem (Markov moves for $B_n^g$
can be defined completely analogous to the usual ones) was only stated
as a conjecture in \refmark{\rSos}.

In \refmark{\rKSS} it was already pointed out that for genus one the
fourth equation of \bralg\ is just RE \re\ if we identify $\s_i = P R
\equiv \Rh $ and $ \t = K_1 $. Knowing now that a solution $K$ of \re\
can be extended to a set of solutions $K^{(\a)}$ having nontrivial
commutation relations \retwi\ we see immediately that \bralg\  is
equivalent to YBE \ybe, RE \re\  and commutation relations \retwi\
if we identify
$$ \eqalign{\s_i &= -q 1 \otimes \ldots \otimes \Rh_{i,i+1} \otimes
\ldots \otimes 1, \quad i=1,\ldots,n-1 \cr
\t_\a &= K^{(\a)} \otimes 1 \ldots \otimes 1 , \quad \a=1,\ldots,g \cr}
\eqn \ident $$
with two more rather obvious consistency conditions as given in \bralg.

Thus $\s_i$ has an explicit matrix representation, but what about
$\t_\a$? Fig.2 suggests to represent the effect of a handle on a
strand carrying a QG representation by $ {K^{(\a)}{}^i}_j =
({{\Rh}^2_{(\a)}}{}^i{}_j)^m{}_n $.
Here $(m,n)$ are the indices of the
QG representation of the first strand `interacting' with the handle,
which is characterized by the special indices $(i,j)$. Therefore we
have a two-dimensional representation of \alg, but for $\{a \leftrightarrow
d, \   b \leftrightarrow c\}$.
To be consistent we adopt the convention to read crossings
involving thick lines corresponding to handles as $\Rth = P \Rh P = R P$
instead of $ \Rh $. So we put $ {K^{(\a)}{}^i}_j
= ({{\Rth}{}^2_{(\a)}}{}^i{}_j)^m{}_n $ and from \rmat\  we calculate
$ {a^m}_n = \pmatrix{q^2 & 0 \cr 0 & 1 \cr} $,
$ {b^m}_n = \pmatrix{0 & 0 \cr \w & 0 \cr} $,
$ {c^m}_n = \pmatrix{0 & \w \cr 0 & 0 \cr} $,
$ {d^m}_n = \pmatrix{1 + \w^2 & 0 \cr 0 & q^2 \cr} $,
which indeed satisfy \alg. It is easy to generalize this to arbitrary
representations the strands may be carrying. In the fundamental
representation we get from \lsol\  just $ S(L^-) L^+ \vert_{\rho{}_{\rm fund}}
= q^{-1} \Rth{}^2 $. Thus we represent $\t_\a$ as
$$ \t_\a = q  S(L^-) L^+ \vert_\rho \otimes 1 \ldots \otimes 1
,  \eqn \taurep  $$
where $ K = q S(L^-) L^+ $ is in an arbitrary representation $\rho$
and the index $\a$ only keeps track of handle numbering. This operator
appeared also in the context of conformal field theory \refmark{\rMoRe}
and was used there, for example, in connection with topology changing
amplitudes in Chern-Simons field theory.

%%%%%%%%%%%%%%%%%%%%%%%%%%  Section 4  %%%%%%%%%%%%%%%%%%%%%%%%%%%%%%

\section{Invariant link polynomials on solid tori}

There are several approaches to the construction of link polynomials,
one may roughly distinguish them in the following way (a convenient
access to literature is \Ref\rKoh{\Koh}). It is well known that the
expression of $\s_i$ in terms of $\Rh$ gives rise to a Hecke algebra
representation of the braid group $B_n$ and the
characteristic equation $ \Rh^2 = \w \Rh + 1 $  of the $R$-matrix
together with the first two equations of \bralg\  comprise just the
relations of the Hecke algebra $H(q^2,n)$ with generators $\s_i$.
One defines a linear functional on $H(q^2,n)$, the Ocneanu trace,
which is the main ingredient in the definition of the invariant link
polynomial \Ref\rJon{\Jon}. Alternatively one may use the matrix trace
of the braid group generators represented by $\Rh$ and then prove
invariance w.r.t Markov moves \REF\rTur{\Tur} \REF\rADW{\ADW}
\refmark{\rTur,\rADW}. Further it is possible to define link
polynomials recursively using skein relations \REF\rAlex{\Alex}
\REF\rCon{\Con} \REF\rKauf{\Kauf} \refmark{\rAlex,\rCon,\rKauf}.
Finally, there is the Chern-Simons field theory approach
\Ref\rWit{\Wit}.

In view of this we may expect that the explicit representation \ident\
and \taurep\  can be used to define an invariant link polynomial on
$H_g$ by means of (quantum) traces of generators $\s_i$ and $\t_a$.
However we shall outline the simpler approach via skein relations here.

Above characteristic equation of the $R$-matrix is equivalent to the
skein relation of the Jones polynomial. It means that the invariant
polynomial satisfies for each oriented link $L$
$$\a P_+(L) +\b P_-(L) + \c P_0(L) = 0 , \quad \a,\b,\c={\rm const.}
\eqn  \skeina  $$
Generally, the invariant linear functional $P(L)$ may depend on one or
more parameters and $P_{\pm},P_0$ are its value for a link $L$ which is
different in each case only at a single crossing as shown in Fig.3:
\vskip1cm
\centerline{\psfig{figure=LINKfig3.ps,width=10cm}}
\centerline{Fig.3: \  Skein relation of type A}
\vskip1cm  \noindent
One may read these crossings as $\Rh,\ \Rh^{-1}$ and $I$ for example.
We keep $\a,\b,\c$ arbitrary as we do not want to specify a certain
polynomial, nor do we fix the type of isotopy (i.e. ambient or regular).
We only assume we are given a skein relation and have normalized the
polynomial of the unknot. This is sufficient to construct the
polynomial of any link uniquely, starting from the unknot, or vice
versa. Our strategy is to use this procedure to unknot any link on
$H_g$ completely which is clearly possible. Then we end up with
unknots which do
\item{(i)} not go around a handle,
\item{(ii)} go around a handle once,
\item{(iii)} go several times around a handle,
\item{(iv)} go around two (or more) handles once (or several times),

\noindent we depict them in Fig.4:
\vskip1cm
\centerline{\psfig{figure=LINKfig4.ps,width=14cm}}
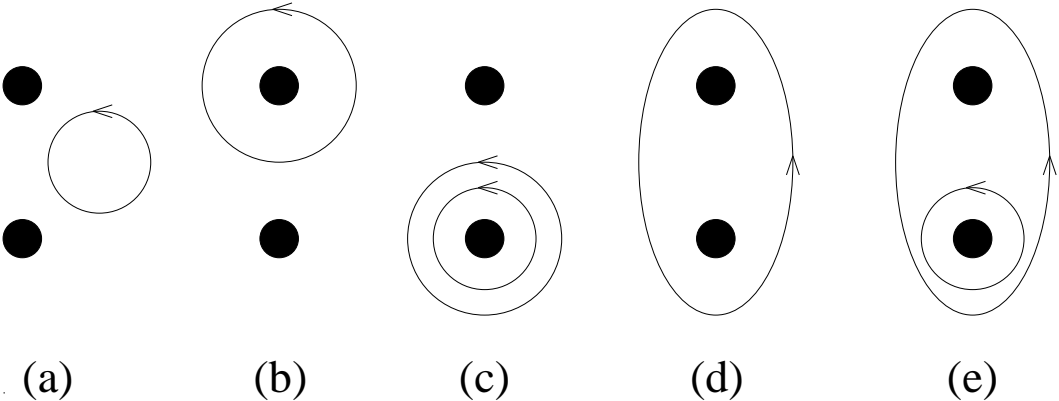
\centerline{Fig.4: \  Some examples of cases (i) - (iv) for genus two}
\vskip1cm  \noindent
Our assertion is that all cases can be reduced to (i) by using \cheq.
This relation can be represented as in Fig.5 and we refer to it as
type B skein relation henceforth.
\vskip1cm
\centerline{\psfig{figure=LINKfig5.ps,width=14cm}}
\centerline{Fig.5: \  Skein relation of type B}
\vskip1cm  \noindent
To unknot any link in $H_g$ we fix the procedure as follows:
\item{1.} Eliminate $\t_a^{-1}$ from the link using type B skein
relations
\item{2.} Unknot the link using type A skein relations
\item{3.} Relate cases (iv), (iii), (ii) $\rightarrow$ (i)

\noindent After step 2 we arrive at cases (i) - (iv) described above.
Step 1 guarantees that we have only links going around handles
counterclockwise since step 2 is orientation preserving.
The reduction from (ii) $\rightarrow$ (i) is easily performed.
We simply  close the braids in Fig.5 and use the type I Reidemeister
move depicted in Fig.6a, where $\d=1$ for ambient isotopy.
\vskip1cm
\centerline{\psfig{figure=LINKfig6.ps,width=12cm}}
\centerline{Fig.6: \  Reidemeister moves of type I}
\vskip1cm  \noindent
We get the result shown in Fig.7, where the first link has different
orientation compared to the others.
\vskip1cm
\centerline{\psfig{figure=LINKfig7.ps,width=12cm}}
\centerline{Fig.7}
\vskip1cm  \noindent
We might pick up a phase factor when reversing the orientation of a
unknot going around a handle. Denoting this as $\e$, and further
taking into account the normalization of the unknot in topologically
trivial regions which we denote by $N$ we finally obtain the result
given in Fig.8:
\vskip1cm
\centerline{\psfig{figure=LINKfig8.ps,width=12cm}}
\centerline{Fig.8}
\vskip0.8cm \noindent
So this gives the desired expression for case (ii)
unknots. It is clear that a case (iii) unknot like the one in Fig.4c
gives the square of the above result (with $c_1 \rightarrow c'_1$).
Then, case (iv) unknots like the one in Fig.4d which corresponds to
the product $\t_1\t_2$ gives the same result as above, however with
$c_1 \rightarrow C_1$, due to RE multiplication properties \prod\
and type B skein relation \comcen. It is then obvious how to treat the
unknot in Fig.4e. We leave all constants unspecified since we do not
want to embark upon explicit calculations here.

%%%%%%%%%%%%%%%%%%%%%%%  Section 5  %%%%%%%%%%%%%%%%%%%%%%%%%%%%%

\section{Summary}

We set out to explain in this paper the braid group on a solid
handlebody and its equivalence to the YB and RE operator algebra.
Further we suggested an explicit representation which could possibly
be used to define invariant link polynomials. We found a way to
construct the polynomial of any link on the handlebody recursively via
new skein relations. These additional skein relations allowed us to
relate the unknot going around a handle to the unknot in topologically
trivial region which is normalized to a constant.

It was not our intention to give rigorous derivations or proofs of the
assertions in section four, rather we would like to draw attention to
the subject, especially it is interesting to see whether this can be
used to construct invariant link polynomials on arbitrary 3-manifolds
as mentioned in the introduction. Then, of course, can this be applied
to conformal and Chern-Simons field theory?
\vskip1cm
{\bf Acknowledgement:}
I would like to express my gratitude to the members of the Yukawa
Institute for Theoretical Physics for extending their hospitality to
me during a rather long stay and for special help when finishing this
manuscript.

\vfill\eject

\refout
\bye